# Understanding algorithmic fairness for clinical prediction in terms of subgroup net benefit and health equity

Jose Benitez-Aurioles[1], Alice Joules[2], Irene Brusini[2], Niels Peek[3], Matthew Sperrin[1]

[1] Centre for Health Informatics, University of Manchester, Manchester, United Kingdom

[2] AI for Healthcare & MedTech, IQVIA, London, United Kingdom

[3] THIS Institute, University of Cambridge, Cambridge, United Kingdom

***Corresponding author***

Jose Benitez-Aurioles, jose.benitez-aurioles@postgrad.manchester.ac.uk, Vaughan House, Portsmouth Street, Manchester, Greater Manchester, M13 9GB.

***Type of manuscript:*** Original research article.

***Suggestion for running head:*** Fairness for clinical prediction using net benefit.

***Conflicts of interest:*** The authors report no conflicts of interest.

***Sources of funding:*** This work was supported by a Health Data Research UK-The Alan Turing Institute Wellcome PhD Programme in Health Data Science studentship (Grant Ref: 218529/Z/19/Z). Funding agencies had no role in the design and conduct of the study.

***Data and computing code availability:*** Data used in this study is sourced from UK Biobank, and was approved with ID 101874. Full code for this work can be found at (https://github.com/jaurioles/FairSubgroupBenefit ).

# Abstract

There are concerns about the fairness of clinical prediction models. 'Fair' models are defined as those for which their performance or predictions are not inappropriately influenced by protected attributes such as ethnicity, gender, or socio-economic status. Researchers have raised concerns that current algorithmic fairness paradigms enforce strict egalitarianism in healthcare, levelling down the performance of models in higher-performing subgroups instead of improving it in lower-performing ones.

We propose assessing the fairness of a prediction model by expanding the concept of net benefit, using it to quantify and compare the clinical impact of a model in different subgroups. We use this to explore how a model distributes benefit across a population, its impact on health inequalities, and its role in the achievement of health equity. We show how resource constraints might introduce necessary trade-offs between health equity and other objectives of healthcare systems.

We showcase our proposed approach with the development of two clinical prediction models: 1) a prognostic type 2 diabetes model used by clinicians to enrol patients into a preventive care lifestyle intervention programme, and 2) a lung cancer screening algorithm used to allocate diagnostic scans across the population. This approach helps modellers better understand if a model upholds health equity by considering its performance in a clinical and social context.

**Word count**: *3932 (approx.)*

# Background

Clinical prediction models are increasingly used to inform diagnostic and prognostic decision-making in healthcare. However, concerns have been raised about these models displaying unequal performance across subgroups defined by protected attributes such as ethnicity, gender or socioeconomic status[1–3]. These disparities can emerge due to poor representation in the training data, unreliable diagnostic labels, or differences in data quality[2,4]. Clinical prediction models thus carry the risk of exacerbating existing health inequalities, driving resources away from populations in need[5]. In the recently released TRIPOD+AI guidelines for the reporting of clinical prediction models, eleven out of the twenty-seven suggested items urge developers to consider the model's impact on underserved populations[6].

Discussions around how researchers should think about the impact of prediction models on underserved populations have not been limited to healthcare[7]. Algorithmic fairness, the study of how models can be made fairer, is an active field of research in machine learning[8], informed by problems in recidivism[9], advertisement[10], and facial recognition[11]. Largely, the focus has been around ensuring that predictions satisfy a specific mathematical definition of fairness for well-defined subgroups[12]. For example, 'demographic parity' states that a prediction model is fair if it, on average, gives the same score to each subgroup. 'Equalised odds' considers a model fair if it has the same false positive and negative rates for all groups[13]. These definitions have been criticised as prioritising an unjustified and excessively egalitarian view of fairness, at least in healthcare. Egalitarianism has been found to, in practice, achieve equality by levelling down the performance of the model in higher performing groups, without increasing it in lower performing ones[14]. While strict equality might be warranted in other domains, it remains inadequate for promoting health equity when developing clinical prediction models[1].

While the concept of algorithmic fairness is relatively new, the unequal access of healthcare by patients has been widely studied in medicine[15–17], and the equitable realisation of health services is a priority in national and international healthcare strategies[18,19]. The goal of health equity is to allow all individuals to achieve their optimal health, irrespective of their sociodemographic status[20] by addressing systematic differences in healthcare opportunities[21,22]. While health equity is an attractive basis for thinking about fairness in clinical prediction models, the challenge of quantifying a model's impact in patients' health remains an obstacle.

In summary, while concerns exist about the equitability of prediction models in healthcare, the topic is methodologically under-developed. In this paper we develop a novel concept of prediction model fairness that is based on the equitable allocation of healthcare resources. We use decision curve analysis[23], a tool to assess the net benefit of prediction models, to quantify and compare resource allocation. Specifically:

1) We will extend the net benefit formula to quantify and compare the clinical utility of a model in different subgroups of a population. We will show how resource constraints create a trade-off between health equity and other healthcare goals, proposing ways to more equitably allocate resources across protected attributes.

2) We will showcase its use by developing two models in different clinical contexts: diabetes preventive management and lung cancer screening, illustrating the insight that the subgroup net benefit can bring to the validation of prediction models in terms of fairness.

## Methods

### *Using subgroup net benefit to investigate model fairness*

Consider a population of size $N$, and a model which gives individual predictions of a binary outcome of interest to inform a clinical decision on this population. A policy is introduced so that individuals with risk scores above $t$ receive one intervention and the rest receive another. The policy can be summarised by its confusion matrix of true positives $TP(t)$, false positives $FP(t)$, false negatives $FN(t)$ and true negatives $TN(t)$. If the utilities $a$, $b$, $c$, and $d$ represent the benefit or cost of each of these outcomes, the overall utility[24] is:

$$U(t) = \frac{1}{N}\big(aTP(t) + bFP(t) + cFN(t) + dTN(t)\big) \qquad (1)$$

The clinically optimal threshold[23] $t^*$ that maximises $U(t)$ can be shown to solve $\frac{1-t^*}{t^*} = \frac{a-c}{d-b}$, so that the utility can be written as:

$$U(t^*) = \frac{a-c}{N}\left(TP(t^*) - \frac{t^*}{1-t^*}FP(t^*)\right) + (c-d)\pi + d \qquad (2)$$

Where $\pi$ is the prevalence of the outcome. Setting without loss of generality the benefit of true positives as unit, so that $a - c = 1$, and ignoring all terms which are model-independent ($(c-d)\pi$ and $d$), the net benefit is derived as:

$$NB(t^*) = \frac{1}{N}\left(TP(t^*) - \frac{t^*}{1-t^*}FP(t^*)\right) \qquad (3)$$

If the optimal threshold $t^*$ is not known, or is not homogeneous across a population, a plot of the net benefit over a range of possible thresholds is plotted, called the decision curve[23].

The term $(c-d)\pi$ was removed as it is model-independent. However, it is not subgroup-independent, making the net benefit of (3) inadequate when comparing subgroups and approaching health equity in the context of pre-existing health inequalities in the population. For example, the net benefit of a policy where no one is treated would be 0 in any subgroup. However, from a health equity perspective, such a policy would more harshly affect groups with high prevalence of the outcome, and thus the impact of the policy is not equal across the population.

We propose reintroducing the prevalence term in the equation. When manipulating $U(t^*)$, linear transformations will not change how models are ranked, therefore we can divide expression (2) by $d - c$:

$$\frac{1}{d-c}U(t^*) = \frac{d}{d-c} - \pi + \frac{1}{N}\frac{a-c}{d-c}\left(TP(t^*) - \frac{t^*}{1-t^*}FP(t^*)\right) \qquad (4)$$

Instead of taking as unit the value of one true positive per capita, so $a - c = 1$, as done for expression (3), we choose as unit the value of one true negative over a false negative, so $d = 1$ and $c = 0$. We thus use as unit the difference between untreated 'healthy' (without outcome) and 'unhealthy' (with outcome) patients. This gives an adapted version of $NB(t^*)$, which we call the subgroup net benefit, $sNB(t^*)$:

$$sNB(t^*) = 1 - \pi + \frac{\lambda}{N}\left(TP(t^*) - \frac{t^*}{1-t^*}FP(t^*)\right) \qquad (5)$$

In this equation, $1 - \pi$, the prevalence of patients without outcome, is the benefit of the group before any model is introduced. A group with $sNB(t^*) = 1$ is fully unburdened by the outcome as it only has true negatives (or 'fully cured' true positives). A group with $sNB(t^*) = 0$ only has false negatives, individuals who will experience the outcome of interest but aren't treated.

The term $\frac{\lambda}{N}\left(TP(t^*) - \frac{t^*}{1-t^*}FP(t^*)\right)$ corresponds to the conventional net benefit, weighted with $\lambda = \frac{a-c}{d-c}$, the relative benefit of treating a true positive $(a - c)$ compared to being a true negative $(d - c)$. A choice of $\lambda = 1$ assumes that the treatment is fully effective in true positives, while $\lambda = 0$ assumes that the treatment is useless. We propose to choose $\lambda$ by considering a surrogate adverse event related to the outcome (e.g., event-related mortality), or the outcome itself in preventive prognosis. We would then use $\lambda = RRR$ where $RRR$ is the relative risk reduction of the treatment allocated by the model on the surrogate event. Different values of $\lambda$ and $t^*$ can be chosen, so that each subgroup $\{g\}_{g=1..G}$ has a different treatment effect $\lambda_g$ and corresponding optimal threshold $t^*_g$. A more detailed derivation and justification for the use of the $RRR$ is provided in Supplementary 1.1.

An important property of the sNB is that it is collapsible, as the aggregated $sNB_{1+2}$ over two groups of sizes $N_1$ and $N_2$, prevalences $\pi_1$ and $\pi_2$, treatment effects $\lambda_1$ and $\lambda_2$, and optimal thresholds $t_1^*$ and $t_2^*$, is the weighted average of their subgroup net benefits $sNB_1$ and $sNB_2$:

$$sNB_{1+2}(t^*) = 1 - \frac{N_1\pi_1 + N_2\pi_2}{N_1 + N_2}$$

$$+ \frac{\lambda_1 N_1}{N_1 + N_2}\left(TP_1(t^*) - \frac{t_1^*}{1 - t_1^*}FP_1(t^*)\right)$$

$$+ \frac{\lambda_2 N_2}{N_1 + N_2}\left(TP_2(t^*) - \frac{t_2^*}{1 - t_2^*}FP_2(t^*)\right)$$

$$= \frac{N_1}{N_1 + N_2}sNB_1(t^*) + \frac{N_2}{N_1 + N_2}sNB_2(t^*) \qquad (6)$$

Where $TP_1(t^*)$, $TP_2(t^*)$, $FP_1(t^*)$ and $FP_2(t^*)$ are the true and false positives in both groups.

*Considering health equity when developing clinical prediction models*

We propose examining the fairness of a model by comparing the sNB across protected attributes of interest, before and after the introduction of the model. The goal of a model is to maximise the net benefit, so that the entire population benefits the most from a policy change. During model development, the goal of health equity is aligned with that of maximising the sNB in the subgroup with the smallest sNB, thus improving the 'maximin' of the model's sNB across groups. More generally, an equitable goal would be to reduce the gap in sNB between groups with high sNB and groups with low sNB by 'levelling up' the latter. Since the sNB is collapsible across subsets of the population, no trade-off is required, provided that there are no resource constraints. To maximise both the overall net benefit and the maximin of sNB across protected attributes, you could choose the model most beneficial in each subgroup and create an ensemble model that is therefore the most beneficial in the overall population.

In the case of resource constraints, not all policies are possible, and policies in different groups are not independent of each other, as allocating more resources to one group entails allocating less resources to another. Because of this, a trade-off between the overall NB and maximin of sNB across protected attributes might be necessary. In that case, the choice of model threshold $t$ cannot always be the clinically optimal threshold $t^*$. Consider a situation in which a maximum capacity is set for the number of positives that the policy is allowed to allocate, making it impossible to choose $t = t^*$. The first potential choice of $t$ would be one $t > t^*$ which maximises the overall net benefit within the capacity constraints of the model. However, allocating comparatively more resources to the most underserved subgroups by making their threshold closer to $t^*$ could improve the maximin sNB. A Pareto front[14,25,26] for this trade-off can be identified, so that we can visualise choices of thresholds for which no other choice exists which has both larger overall net benefit and maximin sNB. Decision makers could examine the Pareto curve, identifying which policy is most appropriate for deployment.

*Showcase of the use of subgroup net benefit in clinical prediction modelling*

To showcase how the sNB can be used, we consider two use cases.

1) *Diabetes prognostic model:* The aim is to build a risk prediction model that will assist clinicians in managing patients at risk of developing type-2 diabetes by referring them to a lifestyle intervention programme that facilitates changes in diet and physical activity. We defined the outcome of interest as 5-year incidence of type-2 diabetes. We choose the optimal clinical threshold for such an intervention to be $t^*_{diabetes} = 15\%$, informed by the use of the Leicester Diabetes Risk Score in clinical practice[27,28]. We chose as a proxy variable the development of diabetes after intervention, as the goal of prognostic modelling and preventive care is to reduce the incidence of the predicted disease. Given this, we choose a treatment weight of $\lambda_{diabetes} = 0.58$, corresponding to the reported RRR of a particular diabetes prevention programme[29]. In the UK, type-2 diabetes prevalence is about three to five times higher in minority ethnic communities[30]. Because of this, we investigated the impact of the developed model in different ethnicities (Black, Asian, white, and other ethnicities).

2) *Lung cancer screening algorithm:* The aim is to build an algorithm that will systematise which individuals get screened, with the aim of detecting more lung cancer cases at an early stage. We defined the outcome of interest as 6-year incidence of lung cancer. We considered a single threshold (corresponding to the allocation of a CT scan) where $t^*_{lungcancer} = 1.5\%$ is chosen as it is the threshold used in screening programmes in the UK[31] and the US[32]. Based on a review of lung cancer screening policies[33], we chose $\lambda_{lungcancer} = 0.20$ as an estimate of the RRR of cancer-related death, its reduction being the main goal of cancer screening efforts. The targeting of deprived areas has been a priority of the lung cancer screening efforts in the UK[34], and we evaluated the changes in sNB by quintiles of the Townsend score (a measure of social and economic deprivation used in the UK). We considered restrictive policies, where only up to 3% and 1% of the population could be allocated screening scans, exploring different choices of thresholds, from $1.5\%$ to $10\%$, across the most-deprived and other quintiles of deprivation.

Cohorts of patients between 40 and 70 years old at baseline were extracted from UK Biobank[35]. Relevant demographic and clinical predictors were identified for each use case (Supplementary Table 1), and a single imputed dataset with multivariate imputation by chained equations[36] was used in order to deal with missing predictor data. Three models were developed for each use case:

1) A logistic regression model which does not include the protected attribute of interest as a predictor. (LogNoSA)

2) A logistic regression model which includes it. (LogSingleSA)

3) An ensemble of logistic regression models, where one model is trained for each group. In order to increase the sample size of each model, data from individuals outside the group are included during training using propensity score weighting[37,38]. This is used to personalise models to target populations of interest (in this case, defined by protected attribute) by borrowing data from individuals of a source population that are most similar to the target group. This should, in theory, reduce statistical bias in the predictions compared to models trained in unweighted data (see Supplementary 1.2 for further information). (LogMultiSA)

The models are trained and validated using performance metrics corrected for optimism using 500 bootstrap samples[39]. The discrimination (through the C-Statistic), calibration (through the slope and intercept) and decision curves of each model in each subgroup are reported. The sNB across the protected attribute of interest is reported for each model, as well as the sNB of a policy in which no patient receives an intervention, considered as the baseline policy for comparison, and a 'purely fair' policy where 5% of individuals are screened, chosen randomly and independently of protected or non-protected attributes. Pareto optimality curves between the overall and maximin sNB are plotted in the resource-constrained lung cancer examples. XGBoost versions of the three models above were also explored, trained and validated using 85:15 split-sampling, and the results for these are presented as supplementary material.

## Results

### Diabetes prognostic risk model

For the development and validation of the models, 477,558 patients were identified, of which 2.1% were Asian, 1.6% were Black, 94.3% were white, 1.8% identified as another ethnicity, and 0.3% had missing ethnicity information. A more detailed table of the cohort is included in Supplementary 2 Table 2.

The performance of the logistic regression models is shown in Figure 1. The calibration intercept showed under-prediction of risk in the Asian, Black and other ethnicity groups for the LogNoSA model compared to the other models, while the calibration slope showed that predicted risks were too extreme in these groups for all models. The sNB is shown in Figure 2. When considering the policy where no one is treated, the sNB of the population was highest in the white group (in TNs per 10,000 patients: 9,801), then the other ethnicity group (9,679), the Black group (9,572), and finally the Asian group (9,380). The model LogSingleSA improved the net benefit in the Asian (9,435), Black (9,597), other (9,690), and white (9,808) populations, although the difference in the latter two groups was very small. It also reduced the health gap between the most benefitted (white) group and the least benefitted (Asian) group by 48 (95% CI: 32 – 66), and narrowed the average gap between the white group and the other three groups by 24 (13 – 33), compared to the policy that considers everyone to be low risk. The model that didn't include ethnicity in their predictors, LogNoSA, benefitted the Asian population less (Difference in the sNB between the models LogNoSA and LogSingleSA: 32; 16 – 49), since, as shown by its calibration intercept, it had underestimated risk scores in these populations, missing more false negatives without reducing enough the number of false positives.

While the policy where a random 5% of patients are screened reduced the gap between the most (Asian) and least (white) affected subgroups by 25, it did so by harming both through false positives. The ensemble of logistic regressions LogMultiSA performed comparably to LogSingleSA. The decision curves of all models for each group are plotted in Supplementary 2 Figure 1, and the results for the XGBoost models are plotted in Supplementary 2 Figures 2, 3 and 4. Overall, the introduction of a clinical prediction model was beneficial for all ethnicities and reduced healthcare inequality gaps.

*Lung cancer screening algorithm*

A cohort of 480,488 patients were identified, with a median Townsend score of -2.14. A more detailed table of the cohort is included in Supplementary 2 Table 3.

The performance of the logistic regression models is shown in Figure 3, with no notable difference in performance across models. The sNB for each quintile of deprivation is shown in Figure 4. Without screening programme, the sNB outside the least deprived quintile was similar across the quintiles (for the least deprived quintile, in TNs per 10,000 patients: 9,975) and was lowest in the most deprived quintile (9,933). There was no meaningful difference in terms of sNB change between the three models. LogSingleSA improved the sNB of all quintiles (in order of deprivation: 9,976; 9,971; 9,970; 9,961; and 9,938) compared to screening no-one. This increase was nonetheless small, with the highest increase being in the second-most deprived (2.1; 95% CI: 1.6 – 2.6) and most deprived (5.0; 4.2 – 5.8) quintiles. The difference between the highest and lowest quintile of deprivation was smaller when implementing any of the three models compared to no screening (38 and 42, respectively). The random screening policy slightly reduced gaps in sNB between quintiles, but did so by harming all groups. The decision curves of all models for each group are plotted in Supplementary 2 Figure 5. Overall, the introduction of a clinical prediction model to the screening pipeline was beneficial for all deprivation quintiles and reduced healthcare inequalities by especially advantaging the most-deprived quintile.

At the clinically optimal threshold of 1.5%, all logistic regression models identified less than 5% of the population for screening (for LogNoSA, LogSingleSA, LogMultiSA respectively: 4.8%, 4.8%, and 4.9%). Investigating stricter resource constraints of only screening 3% and 1% of the eligible population, a trade-off between overall net benefit and sNB in the highest deprivation quintile is identified through a Pareto front (Figure 5). The differences in overall benefit were very small across the Pareto front (for LogSingleSA, the difference between the highest and smallest overall sNB is 0.8 TNs per 10,000 patients when imposing a 3% cap on the number flagged positive, and 0.7 when imposing a 1% cap). The difference in the sNB of the highest deprivation quintile across the Pareto frontier was relatively large (4.2 when imposing a 3% cap, 2.8 when imposing a 1% cap). The results for the XGBoost model are plotted in Supplementary Figures 6, 7, 8, and 9. The performance of the XGBoost model was similar to that of the logistic regression model, without important differences between the two.

Discussion

In this paper, we presented an extension of net benefit to better understand the role of a clinical prediction model in upholding health equity. We showcased this with two examples in which the subgroup net benefit brought insight around the fairness of the models being developed. In the considered examples, there exist large health inequalities in the population, as shown through gaps in the subgroup net benefit between groups. Introducing clinical prediction model-informed policies did not completely solve this, but is better than alternative policies. In the type-2 diabetes prognosis example, modelling choices mattered, with models accounting for ethnicity benefitting the Asian and Black subgroups, and while calibration metrics already suggested that accounting for ethnicity would improve performance in these groups, the subgroup net benefit showed how this decision specifically reduced healthcare inequalities, and to what degree. In the lung cancer screening case, all models performed equally well, reducing overall gaps in net benefit between the top quintile of deprivation and the rest of the population. Only very restrictive resource constraints needed a trade-off between the net benefit of the overall population and that of the most deprived subgroup.

Some researchers have already suggested alternative algorithmic fairness paradigms where fairness is approached as a multi-objective optimisation problem, where model development aims to jointly maximise overall and per-group performance[14,25,26]. However, choosing an appropriate measure to represent performance can be difficult, and ensuring fairness according to one metric can be incompatible to ensuring it with another[40]. The net benefit has been used before to investigate model fairness[41], but due to it not being adapted for subgroup comparisons[42], the focus of this work was on how to separately optimise predictions in each subgroup, instead of comparing the benefit across the population, or considering resource constraints.

Our approach identifies how restricting resources raises a subjective consideration of whether it is more important to allocate our resources to improve the overall health of the population, or to help those most underserved. Paulus & Kent[1] identifies the existence of non-polar predictions, for which the interests of the patient and healthcare system align, and polar predictions, for which they don't. We link this idea to the concept of distributive justice, a key consideration in health equity[20]. Instead of choosing a solution across the Pareto curve, additional resources could be allocated to the problem, in order to alleviate the necessary trade-off. While the calculation of the net benefit under resource constraints has previously been discussed[43], this is, to our knowledge, the first application for fairness concerns and multi-objective optimisation.

In our examples, including protected attributes within models tended to benefit non-majority subgroups. This result has been found in other works that compared models that leveraged protected attributes against those that didn't[44,45]. This has justified the use of predictors like ethnicity or gender in clinical prediction[46,47]. Other researchers have instead argued against the use of such attributes, particularly ethnicity, citing the essentialisation of race and potential bias as issues[48,49]. The performance of the XGBoost models was similar to that of the logistic regression models, and also benefitted from having ethnicity as a predictor. There have been recent concerns of machine learning models being able to accurately predict ethnicity without having access to it, potentially using it as an implicit predictor for other tasks[50]. Our results suggest that this was not the case in the two selected example use cases.

This study had a limited sample size of particular subgroups within the population, lowering the precision of subgroup net benefit estimates. The quantification of uncertainty around whether the model narrows or widens health inequalities could be leveraged in order to determine whether more representative data is needed[51,52]. Small samples also hinder our ability to investigate the subgroup net benefit across all possible groupings that can be made with protected attributes (including intersections of multiple protected attributes), as sampling variability limits the inferences that can be made from the analysis, as is true of any fairness analysis. We recommend, in line with TRIPOD+AI guidelines[6], to focus on divisions (potentially defined by multiple protected attributes) of particular interest given known historical health inequalities.

Our approach requires assumptions around the optimal threshold and the benefit of true positives, the former being a limitation of all decision curve analysis approaches. However, choosing reasonable values for these is relatively simple, and ranges of values can be explored. Although the proposed method does not account for potential unfairness due to differential measurement error or generalisability issues, it can be interpreted in light of any concerns in order to make fairer decisions, enriching a more holistic view of health equity in clinical prediction model development[53].

Unfairness and health inequalities are becoming a growing concern in the development and deployment of clinically useful healthcare algorithms. We recommend that the subgroup net benefit is reported during the development and validation of clinical prediction models. In addition, we recommend checking whether the inclusion of protected attributes, when feasible, as predictors, narrows health inequalities. Future work could use the subgroup net benefit in a practical case, leverage its uncertainty to identify the need for more representative data, or investigate the consequences of assumptions around the threshold and true positive benefit being untrue.

## Acknowledgement

JBA is the receipt of the studentship awards from the Health Data Research UK-The Alan Turing Institute Wellcome PhD Programme in Health Data Science (Grant Ref: 218529/Z/19/Z). We acknowledge support of the UKRI AI programme, and the Engineering and Physical Sciences Research Council, for CHAI - Causality in Healthcare AI Hub [grant number EP/Y028856/1]. The research was carried out at the National Institute for Health and Care Research (NIHR) Manchester Biomedical Research Centre (BRC) (NIHR203308).

This research has been conducted using the UK Biobank resource under application number 101874.

**Figure 1:** Performance in 5-year T2 diabetes risk prediction in different ethnicities and overall, for three models: a logistic regression model which does not use ethnicity as a predictor (LogNoSA), a logistic regression model which does include it (LogSingleSA) and an ethnicity-personalised ensemble of logistic regression models (LogMultiSA). The C-statistic (a), calibration slope (b) and calibration intercept (c) are plotted with 95% confidence intervals.

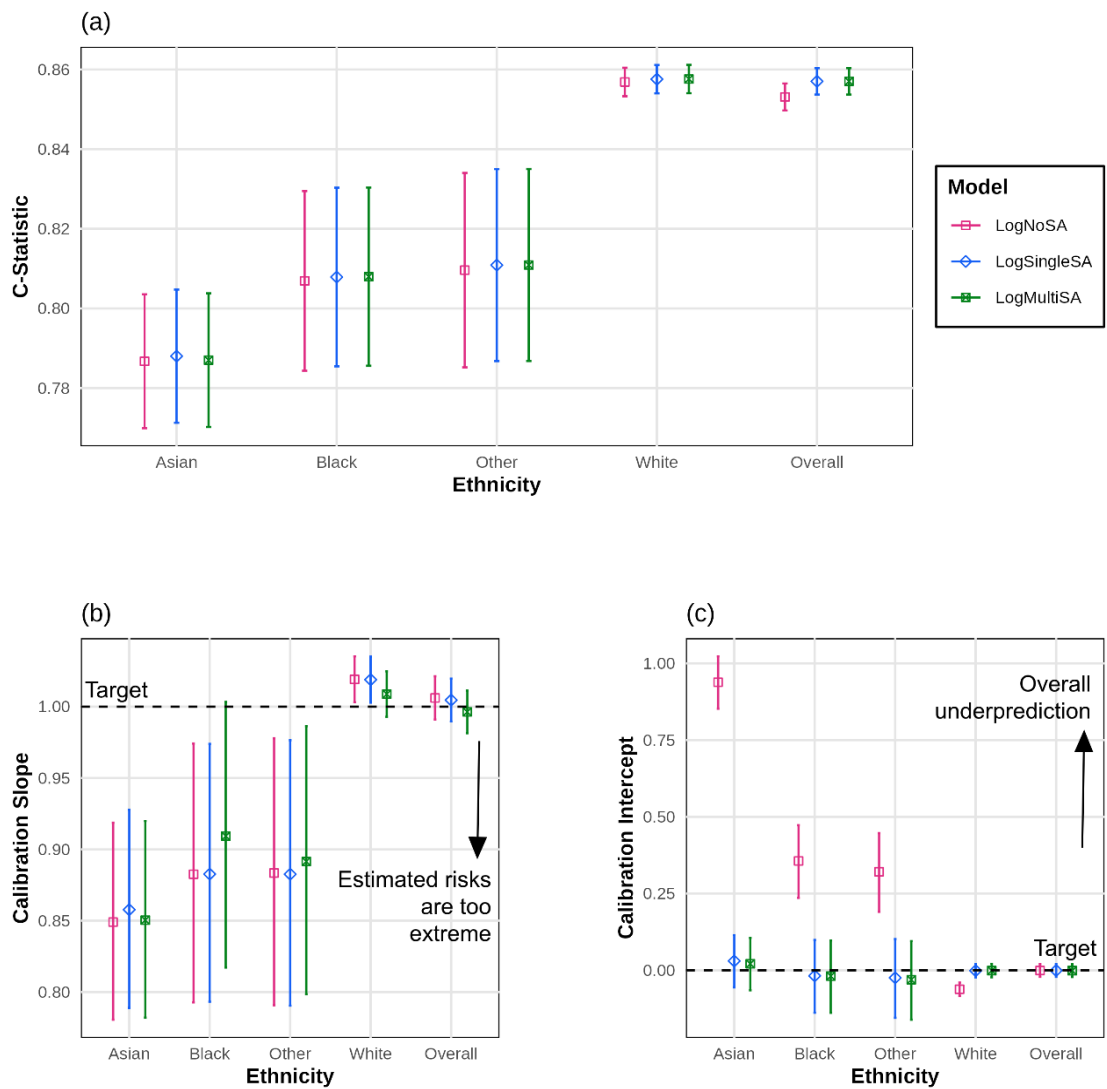

**Figure 2:** Subgroup net benefit, in true negatives per 10,000 patients, in 5-year T2 diabetes risk prediction in different ethnicities and overall, for three models: a logistic regression model which does not use ethnicity as a predictor (LogNoSA), a logistic regression model which does include it (LogSingleSA) and an ethnicity-personalised ensemble of logistic regression models (LogMultiSA). A policy of screening no patients (Treat.No.One), and a policy where a random 5% of patients are screened (Random 5%),  are also plotted.

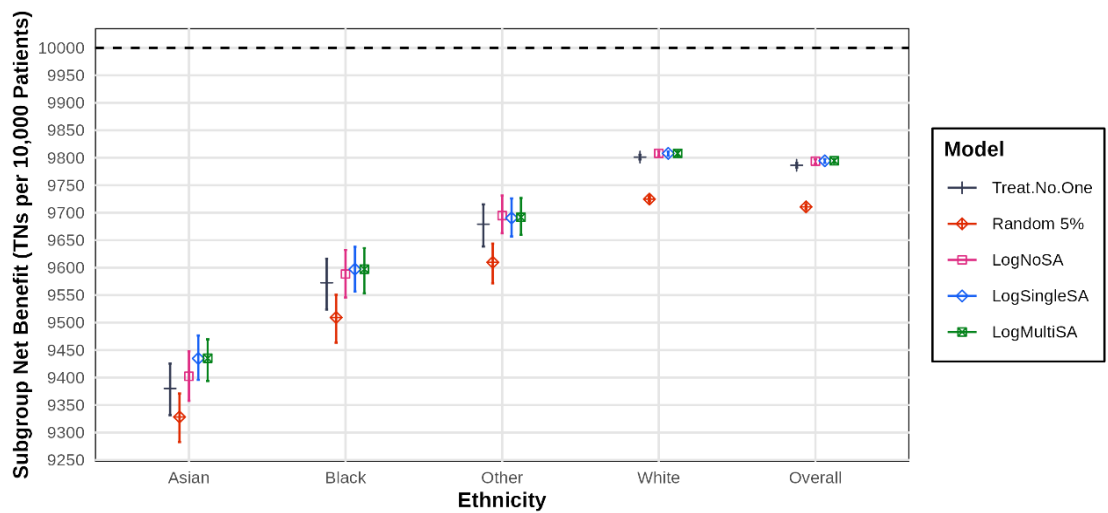

**Figure 3:** Performance in 6-year lung cancer screening for different quintiles of Townsend score and overall, for three models: a logistic regression model which does not use the Townsend score as a predictor (LogNoSA), a logistic regression model which does include it (LogSingleSA) and a quintile-personalised ensemble of logistic regression models (LogMultiSA). The C-statistic (a), calibration slope (b) and calibration intercept (c) are plotted with 95% confidence intervals. "Q5" corresponds the most deprived quintile of the population.

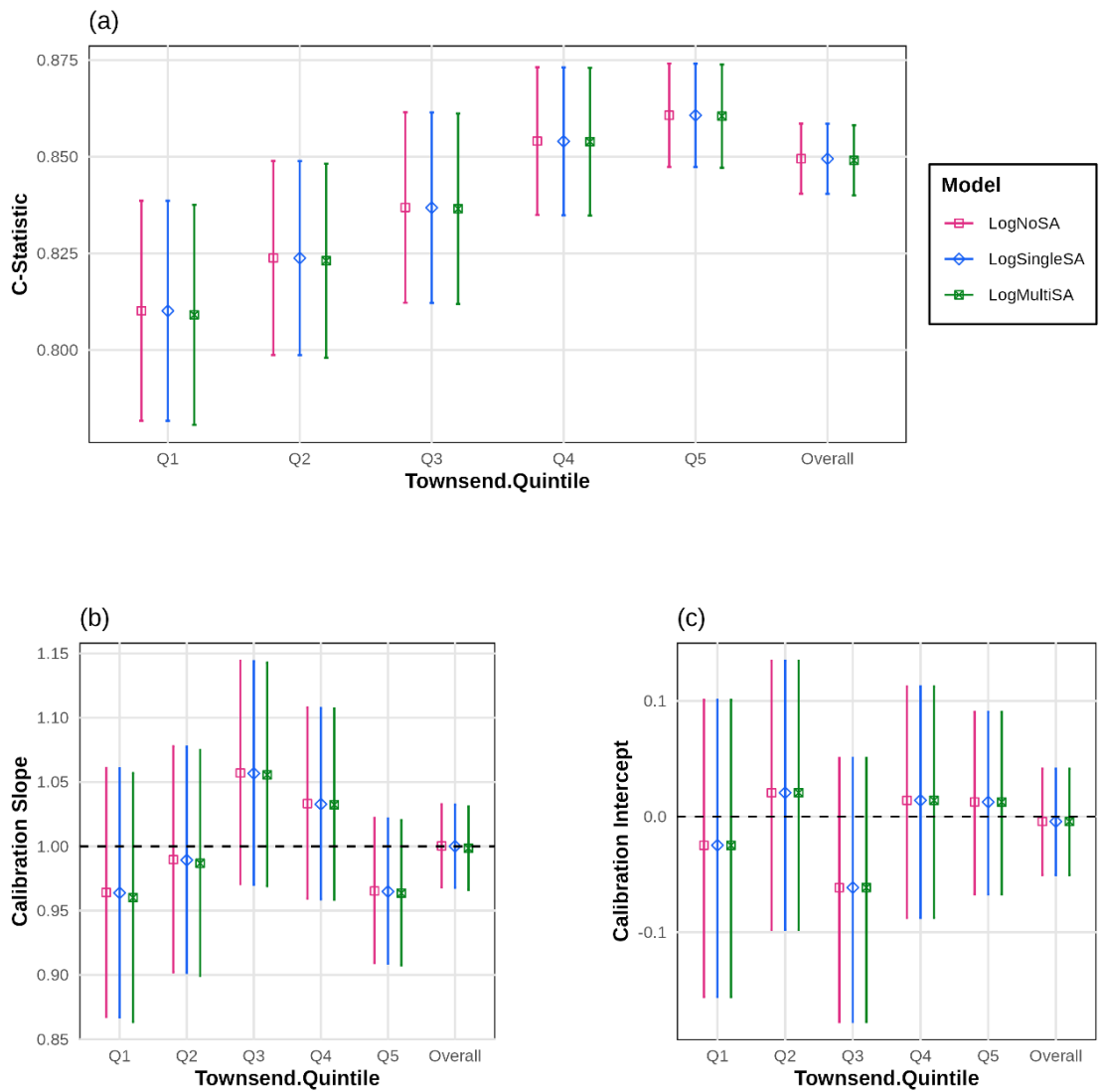

**Figure 4:** Subgroup net benefit, in true negatives per 10,000 patients, in 6-year lung cancer screening for different quintiles of Townsend score and overall, for three models: a logistic regression model which does not use Townsend score as a predictor (LogNoSA), a logistic regression model which does include it (LogSingleSA) and a quintile-personalised ensemble of logistic regression models (LogMultiSA). A policy of screening no patients (Treat.No.One), and a policy where a random 5% of patients are screened (Random 5%), are also plotted. "Q5" corresponds the most deprived quintile of the population.

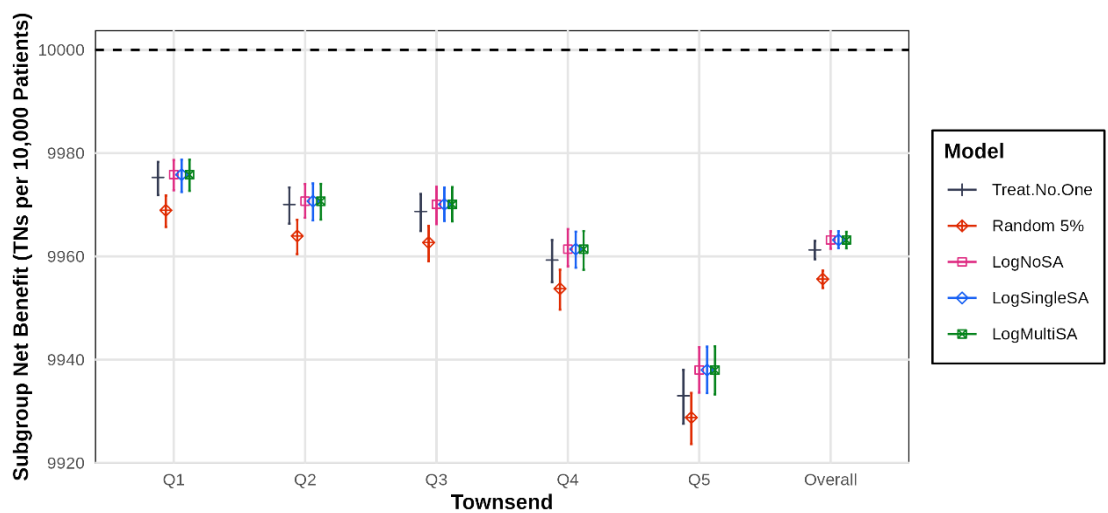

**Figure 5:** Trade-off between the subgroup net benefit in the most deprived quintile of the population and overall, shown through the identified set of thresholds that are Pareto optimal, for three models, in the lung cancer prediction use case: a logistic regression model which does not use Townsend score as a predictor (LogNoSA), a logistic regression model which does include it (LogSingleSA) and a quintile-personalised ensemble of logistic regression models (LogMultiSA). The Pareto front was corrected for in-sample optimism. The policy of screening no patients (Treat No One) is also shown.

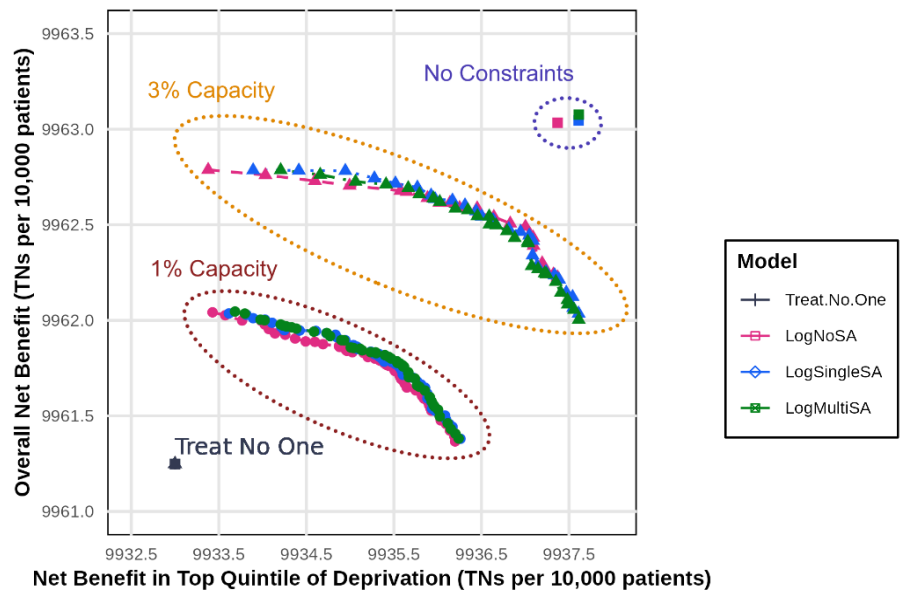

## Supplementary 1.1: Subgroup net benefit derivation

To derive the formula for the subgroup net benefit, we first consider the original conditional utility formula for the utility of a clinical prediction model trying to predict an outcome $Y = 1$ at a particular threshold $t$.

$$U(t) = \frac{1}{N}\big(aTP(t) + bFP(t) + cFN(t) + dTN(t)\big) \quad (S1.1)$$

Where $N$ is the number of individuals in the population, $TP(t)$, $FP(t)$, $FN(t)$ and $TN(t)$ are the true positives, false positives, false negatives and true negatives of the model at threshold $t$, and $a$, $b$, $c$ and $d$ are the corresponding utility weights of the confusion matrix. The threshold at which it would be optimal to use the model, $t^*$, is the probability at which the benefit and harm of treating a patient ($a$ and $b$) completely balances out with the benefit and harm of not treating them ($d$ and $c$). It would thus be better to give treatment to patients above this risk (as the chance of a true positive increases and that of a false positive decreases), while it would be better not to treat a patient below this risk (as the chance of a true negative increases and that of a false negative decreases). The optimal threshold is obtained by considering this balance:

$$at^* + b(1 - t^*) = ct^* + d(1 - t^*) \quad (S1.2)$$

From this expression, we can write the optimal threshold $t^*$ as $\frac{t^*}{1-t^*} = \frac{d-b}{a-c}$. Going back to the utility formula (A1.1), we re-write the false negatives and true negatives as $FN(t) = Pos - TP(t)$ and $TN(t) = N - Pos - FP(t)$ where $Pos$ is the number of positive cases in the data, so that $Pos = TP(t) + FN(t)$ and $Pos = FP(t) + TN(t)$ for any $t$. The conditional utility at the optimal threshold $t^*$ can thus be written as:

$$U(t^*) = \frac{1}{N}\big(aTP(t^*) + bFP(t^*) + cPos - cTP(t^*) + dN - dPos - dFP(t^*)\big)$$

$$= d + (c - d)\,\pi + \frac{1}{N}(a - c)\left(TP(t^*) + \frac{b - d}{a - c}FP(t^*)\right) \quad (S1.3)$$

Where $\pi = \frac{Pos}{N}$ is the prevalence of the outcome in the population. When manipulating the utility, linear transformations (multiplying by, adding, or subtracting any positive constant that does not depend on model or subgroup) will not change how models are ranked, so that if a model or population has a higher utility than another before the transformation, it still has a higher utility after the transformation. In that sense, linear transformations only constitute a change in the unit of the metric. Because of this, we can set, without loss of generality, $d = 1$ as an offset and choose as unit the benefit of a not having the outcome compared to having the outcome when untreated (so $d - c = 1$), and rewrite $\frac{b-d}{a-c}$ as $-\frac{t^*}{1-t^*}$, we get:

$$sNB = 1 - \pi + \frac{\lambda}{N}\left(TP(t^*) - \frac{t^*}{1-t^*}FP(t^*)\right) \quad (S1.4)$$

Where $\lambda = \frac{a-c}{d-c}$ is the ratio between the benefit of being a true positive compared to being a false negative, and the benefit of being a true negative compared to being a false negative. This ratio thus represents the comparative utility of being treated compared to never having the outcome in the first place.

In order to find a reasonable choice for $\lambda$, we consider a situation in which there is a relevant proxy outcome of interest, $R$, which is clinically harmful, for which there is a risk in individuals with the outcome $Y = 1$ that gets lowered if they get treated. For prognostic modelling with preventive goals, this can be the outcome $Y$ itself, so that we consider the relative risks of developing diabetes, for example, before and after preventive care. For an example like diagnostic screening, this can be a proxy like lung cancer-related death in 5 years, which has a risk reduced when screened (and caught early) compared to not screening.

We thus consider the probability of experiencing the proxy outcome for true positives and false negatives, $P(R = 1|TP)$ and $P(R = 1|FN)$. Importantly, we consider that the risk of experiencing the proxy outcome as a true negative is null, so $P(R = 1|TN) = 0$. This holds true for outcomes like lung cancer-related death or diabetes, as it cannot happen in true negatives. Otherwise, approaches like considering 'excess' deaths in true positives compared to false and true negatives (for which this excess would be null) could be considered. Using $R$ as a useful proxy for the utility weights, we write:

$$a = 1 - P(R = 1|TP) \quad (S1.5)$$

$$b = d - \frac{(a - c)\, t^*}{1 - t^*} \quad (S1.6)$$

$$c = 1 - P(R = 1|FN) \quad (S1.7)$$

$$d = 1 \quad (S1.8)$$

Importantly, we're not necessarily considering that the outcome $R$ happens in false positives, but that there is a harm $b$ associated with false positives which is related to the harm associated to $R$, according to $b = d - \frac{(a-c)t^*}{1-t^*}$. Feeding this to equation $(S1.3)$ we obtain:

$$U(t^*) = 1 - P(R = 1|FN)p$$

$$+ \frac{1}{N}\big(P(R = 1|FP) - P(R = 1|TP)\big)\left(TP(t^*) - \frac{t^*}{1 - t^*}FP(t^*)\right) \quad (S1.9)$$

Importantly, we need the net benefit to keep the same ordering as the conditional utility across subgroups and models, so that if one utility is greater than another, then its net benefit is also bigger. As long as the transformation we apply to the utility is monotonic, like that of multiplying by a positive number and adding another, we can still consider the result as the net benefit of the model in a subgroup. We divide $U(t^*)$ by $P(R = 1|FN)$ (a positive constant) and reset the offset term to 1 to achieve:

$$sNB(t^*) = 1 - p + \frac{1}{N}\frac{P(R = 1|FN) - P(R = 1|TP)}{P(R = 1|FN)}\left(TP(t^*) - \frac{t^*}{1 - t^*}FP(t^*)\right) \quad (S1.10)$$

So that $\lambda = \frac{P(R = 1|FN) - P(R = 1|TP)}{P(R = 1|FN)} = RRR$ is the relative risk reduction of the proxy outcome due to treatment in positive patients. A reasonable choice of $\lambda$ can therefore be made by considering a relevant proxy relative risk reduction. Importantly, by doing this, we are not assuming that the utility is solely dependent on this outcome, but rather that the relative benefit of treatment in all proxy outcomes that feed into the utility are on average the same as the relative risk reduction of the proxy outcome of choice.

## Supplementary 1.2: Reducing statistical bias when dealing with subgroup heterogeneity using an ensemble of propensity-weighted models

This method was inspired by approaches in the generalisability literature, and is an application Steingrimsson et al. (https://doi.org/10.1093/aje/kwac128) to subgroup heterogeneity instead of transportability. Consider a population with predictors $X$, an outcome $Y = \{0,1\}$, and a sensitive attribute $A = \{0,1\}$ that defines two subgroups. We accept that there is some heterogeneity between the two subgroups of the population $P(X,Y|A = 0) \neq P(X,Y|A = 1)$. For now, we assume that the relationship between $X$ and $Y$ across both populations is the same, so that $P(Y|X, A = 0) = P(Y|X, A = 1)$, so that the difference in the two distributions relates to a shift in the predictors. We have a corresponding dataset of $N$ patients $\{x_i, y_i, a_i\}_{i=1..N}$.

We now set to train a prediction model which we want to tailor only to group $A = 1$, so that we can consider $A = 1$ our 'target' subgroup and $A = 0$ our 'source' subgroup. The first option is to train a model in the entire dataset $\{x_i, y_i, a_i\}_{i=1..N}$. However, if the model is misspecified (or if, like machine learning methods, it uses regularisation techniques to prevent overfitting), this model will be biased as it is not trained in the target distribution $P(X,Y,A = 1)$, but in the distribution $P(X,Y,A)$. Another option is to only train the model in the subset of the data belonging to the target subgroup, $\{x_j, y_j, a_j = 1\}_{j=1..N_{A=1}}$, of limited size $N_{A=1}$. This will increase the variance of the model, as it is trained on a smaller set of individuals. A third approach is to use data from both $A = 0$ and $A = 1$, but weighting the data during fitting with weights $\{w_i\}_{i=1..N}$ so that individuals with $A = 0$ which are more similar to those with $A = 1$ are weighted more than those that are dissimilar. As shown by Steingrimsson et al. (https://doi.org/10.1093/aje/kwac128), if only using data from the source $\{x_k, y_k, a_k = 0\}_{k=1..N_{A=0}}$, in order to minimise the Kullback-Leibler divergence between the estimated and true $P(Y|X, A = 1)$, we fit the model through a weighted maximum likelihood estimator with weights:

$$w_k = \frac{\Pr(A = 1 | X = x_k)}{\Pr(A = 0 | X = x_k)} = \frac{\Pr(A = 1 | X = x_k)}{1 - \Pr(A = 1 | X = x_k)} \qquad (S2.1)$$

This is, at least asymptotically, the best weighting to choose for training the model, and has been shown to work in transportability problems. The probability $\Pr(A = 1 | X = x_k)$ is the probability that an individual belongs to the target subgroup given their predictors, which can be estimated through a propensity score model

Given that to train our model for subgroup $A = 1$, we have access to both 'source' and 'target' individuals, and choose an alternative set of weights:

$$w_i = \min \left( \frac{\Pr(A = 0)}{\Pr(A = 1)} \frac{\Pr(A = 1 | X = x_i)}{1 - \Pr(A = 1 | X = x_i)}, 1 \right) \quad \begin{array}{l} if\ a_i = 0 \\ \\ if\ a_i = 1 \end{array} \qquad (S2.2)$$

In this case, individuals from the target subgroup $A = 1$ aren't re-weighted, as they already belong to the target distribution $P(X, Y, A = 1)$. Individuals from the source subgroup $A = 0$ are weighted as before according to their odds $\frac{\Pr(A=1|X=x_i)}{1-\Pr(A=1|X=x_i)}$, with a scaling factor $\frac{\Pr(A=0)}{\Pr(A=1)}$ which was previously ignored in (1) as it is a constant, but needs to be kept when combining weighted and unweighted data (as follows from the original derivation of Shimodaira, [https://doi.org/10.1016/S0378-3758(00)00115-4](https://doi.org/10.1016/S0378-3758(00)00115-4)). Finally, the weights of the source subgroup $A = 0$ are capped at 1, so that no individual from outside the subgroup of interest is weighted more than those within the subgroup of interest.

We thus propose to, for each subgroup of interest within the population, train a separate model that is optimised to best perform in the 'target' subgroup. In terms of specific steps, we propose the following:

1) Train a propensity score model, a logistic regression model with LASSO regularisation, to estimate each individual's probability of belonging to the target subgroup, $\Pr(A = 1 | X = x_i)$.

2) Use the propensity score model to generate weights $w_i$ for each individual, according to formula (S1.2).

3) Train the clinical prediction model using maximum likelihood minimisation, weighted with $w_i$.

4) Repeat this for each subgroup of the population in order to create an ensemble of propensity-weighted models.

Validation metrics are calculated as usual, without weighting. When correcting for optimism with bootstrapping for the logistic regression models, we only train the propensity score models once in the original dataset, meaning that we are expecting to underestimate the optimism of the model. Ideally, the propensity score models should be re-trained within each bootstrap of the dataset. This latter approach was unfeasible computationally, and we expect the bias in optimism to be very small, as the study size was large, the propensity score model only had access to predictors but not the outcome, and the final results of this approach (LogMultiSA) weren't better than those of a regularly trained logistic regression model (LogSingleSA).

In the case of the XGBoost model training, because we use split-sampling validation instead of the more computationally expensive bootstrapping, we additionally included a 'forgetting factor' $f$ to the choice of weights:

$$w_i = \min \left( f \times \frac{\Pr(A=0)}{\Pr(A=1)} \frac{\Pr(A=1|X=x_i)}{1-\Pr(A=1|X=x_i)}, 1 \right) \quad \begin{array}{l} if\ a_i = 0 \\[6pt] if\ a_i = 1 \end{array} \qquad (S2.3)$$

This factor allows to down-weight observations from $A=0$, in situations where concept shift (i.e., shifts where $P(Y=1|X, A=0)$ and $P(Y=1|X, A=1)$ are not the same) could potentially make individuals from the 'source' group $A=0$ less like those of the 'target' group $A=1$ than determined by the propensity weights. The factor $f$ is chosen, along with other hyperparameters of the XGBoost model, using cross-validation in the training data.

**Supplementary Table 1:** Candidate predictors for the two developed clinical prediction models.

| PROGNOSTIC DIABETES RISK PREDICTION | SCREENING LUNG CANCER RISK PREDICTION |
|---|---|
| Age, Sex, Living in urban area, Townsend score, Ethnicity, Family history (Hip fracture, Depression, Diabetes, High blood pressure, Cardiovascular disease), Diastolic and Systolic blood pressure, BMI, Percentage of body fat, Waist circumference, Education, Smoking status, Pack-years of smoking, alcohol use, Hysterectomy, Menopause, Current medications (Cholesterol, Blood pressure, Hormone replacement therapy, Contraceptives), Clinical history (Psoriatic arthritis, Juvenile arthritis, Gestational diabetes, Preterm Labour, Arthrosis, Ovarian dysfunction, Schizophrenia, schizotypal and delusional disorders, Systemic connective tissue disorders, Mood disorders, Bipolar disorder or mania, Sleep disorder, Obesity, Cardiovascular and ischaemic disease, Hypertensive disease, Hypercholesterolemia, Other metabolic disorders, Gingivitis and periodontal diseases, Learning disability) | Age, Sex, Living in urban area, Townsend score, Ethnicity, Family history (Hip fracture, Depression, Diabetes, High blood pressure, Cardiovascular disease), Diastolic and Systolic blood pressure, BMI, Percentage of body fat, Waist circumference, Education, Smoking status, Pack-years of smoking, alcohol use, Hysterectomy, Menopause, Current medications (Cholesterol, Blood pressure, Hormone replacement therapy, Contraceptives), Clinical history (Asthma, Chronic obstructive pulmonary disease, Pneumonia, Respiratory tuberculosis, Malignant cancer) |

**Supplementary Table 2:** Demographic characteristics of the cohort identified for the T2 diabetes prediction problem, stratified by ethnicity. Ethnicity information was missing for 1,202 individuals, not appearing on this table. For continuous values, the median and interquartile range are reported, while for categorical values, the count and percentage are reported.

| Characteristic | Asian, N = 10,303 (2.2%) | Black, N = 7,476 (1.6%) | Other, N = 8,363 (1.8%) | White, N = 450,214 (94.5%) |
|---|---|---|---|---|
| **Age** | 52 (46, 59) | 50 (45, 57) | 52 (46, 60) | 58 (50, 63) |
| **Female** | 5,159 (50%) | 4,332 (58%) | 4,698 (56%) | 248,543 (55%) |
| **Townsend Score** | 0.07 (-2.37, 2.42) | 2.84 (0.01, 5.55) | 0.53 (-2.38, 3.72) | -2.28 (-3.71, 0.21) |
| Unknown | 19 | 18 | 17 | 543 |
| **Education** | | | | |
| None Reported | 1,562 (18%) | 972 (17%) | 1,055 (15%) | 75,534 (19%) |
| GCSE-Equivalent | 2,066 (24%) | 1,900 (32%) | 1,532 (22%) | 121,601 (31%) |
| A Level-Equivalent | 884 (10%) | 513 (8.7%) | 824 (12%) | 51,115 (13%) |
| University-Equivalent | 4,243 (48%) | 2,498 (42%) | 3,596 (51%) | 146,067 (37%) |
| Unknown | 1,548 | 1,593 | 1,356 | 55,897 |
| **Hip Fracture in Family** | 4,769 (46%) | 2,952 (39%) | 2,571 (31%) | 89,183 (20%) |
| **Diabetes in Family** | 567 (5.5%) | 228 (3.0%) | 640 (7.7%) | 57,319 (13%) |
| **CVD in Family** | 9,352 (91%) | 6,644 (89%) | 7,196 (86%) | 418,412 (93%) |
| **BMI** | 26.0 (23.7, 28.9) | 28.6 (25.7, 32.1) | 26.8 (24.1, 30.1) | 26.6 (24.1, 29.7) |
| Unknown | 130 | 123 | 133 | 1,745 |
| **Taking Statins** | 2,015 (20%) | 948 (13%) | 1,123 (13%) | 68,347 (15%) |
| **Taking Blood Pressure Medication** | 2,143 (21%) | 2,107 (28%) | 1,429 (17%) | 85,163 (19%) |
| **Taking HRT** | 216 (2.1%) | 177 (2.4%) | 291 (3.5%) | 18,473 (4.1%) |
| **Gestational Diabetes** | 44 (0.4%) | 16 (0.2%) | 21 (0.3%) | 468 (0.1%) |
| **Delusional Disorder** | 33 (0.3%) | 58 (0.8%) | 54 (0.6%) | 1,082 (0.2%) |
| **Manic Episode/Bipolar Disorder** | 22 (0.2%) | 31 (0.4%) | 51 (0.6%) | 1,584 (0.4%) |
| **Hypertension** | 2,616 (25%) | 2,539 (34%) | 1,929 (23%) | 113,188 (25%) |
| **5-Year Diabetes Incidence** | 640 (6.2%) | 320 (4.3%) | 270 (3.2%) | 8,933 (2.0%) |

**Supplementary Figure 1:** Net benefit (in true positives per 10,000 patients) across thresholds for the overall (a), Asian (b), Black (c), other ethnicity (d), and white (e) groups for three models: a logistic regression model which does not use ethnicity as a predictor (LogNoSA), a logistic regression model which does include it (LogSingleSA) and an ethnicity-personalised ensemble of logistic regression models (LogMultiSA). The net benefit of treating no patients (Treat.No.One) and all patients (Treat.All) are also plotted. The chosen optimal threshold to calculate the subgroup net benefit (15%) is shown through a vertical orange line.

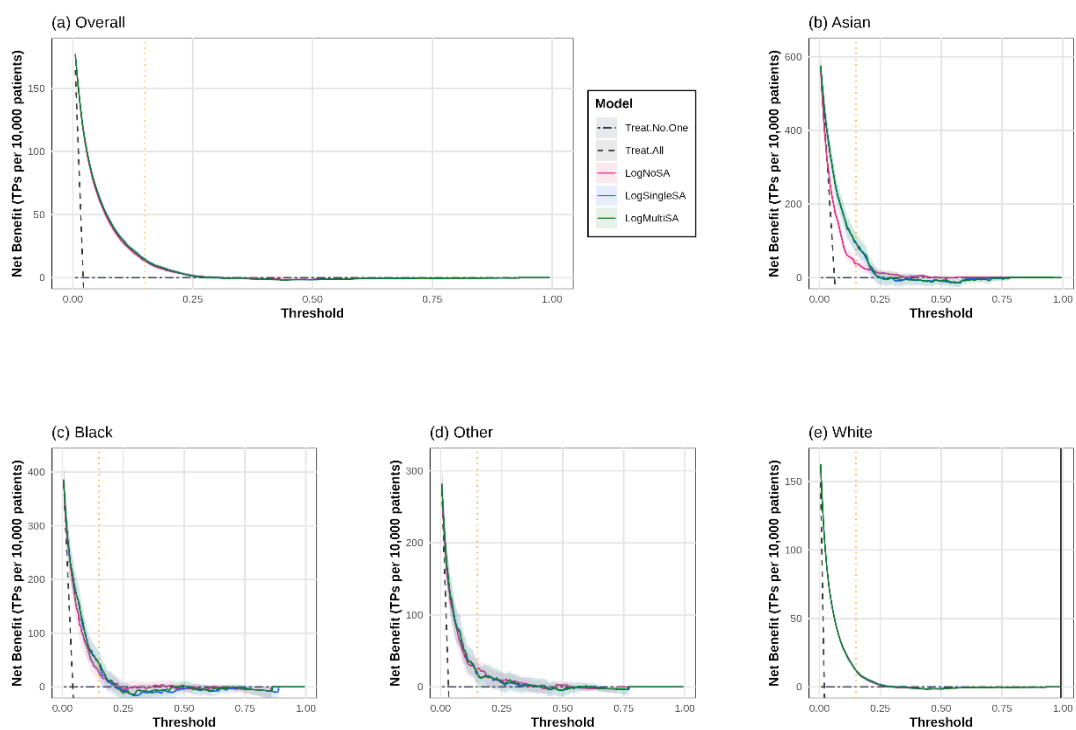

**Supplementary Figure 2:** Performance in 5-year T2 diabetes risk prediction in different ethnicities and overall for three models: a XGBoost model which does not use ethnicity as a predictor (XGBNoSA), a XGBoost model which does include it (XGBSingleSA) and an ethnicity-personalised ensemble of XGBoost models (XGBMultiSA). The C-statistic (a), calibration slope (b) and calibration intercept (c) are plotted with 95% confidence intervals.

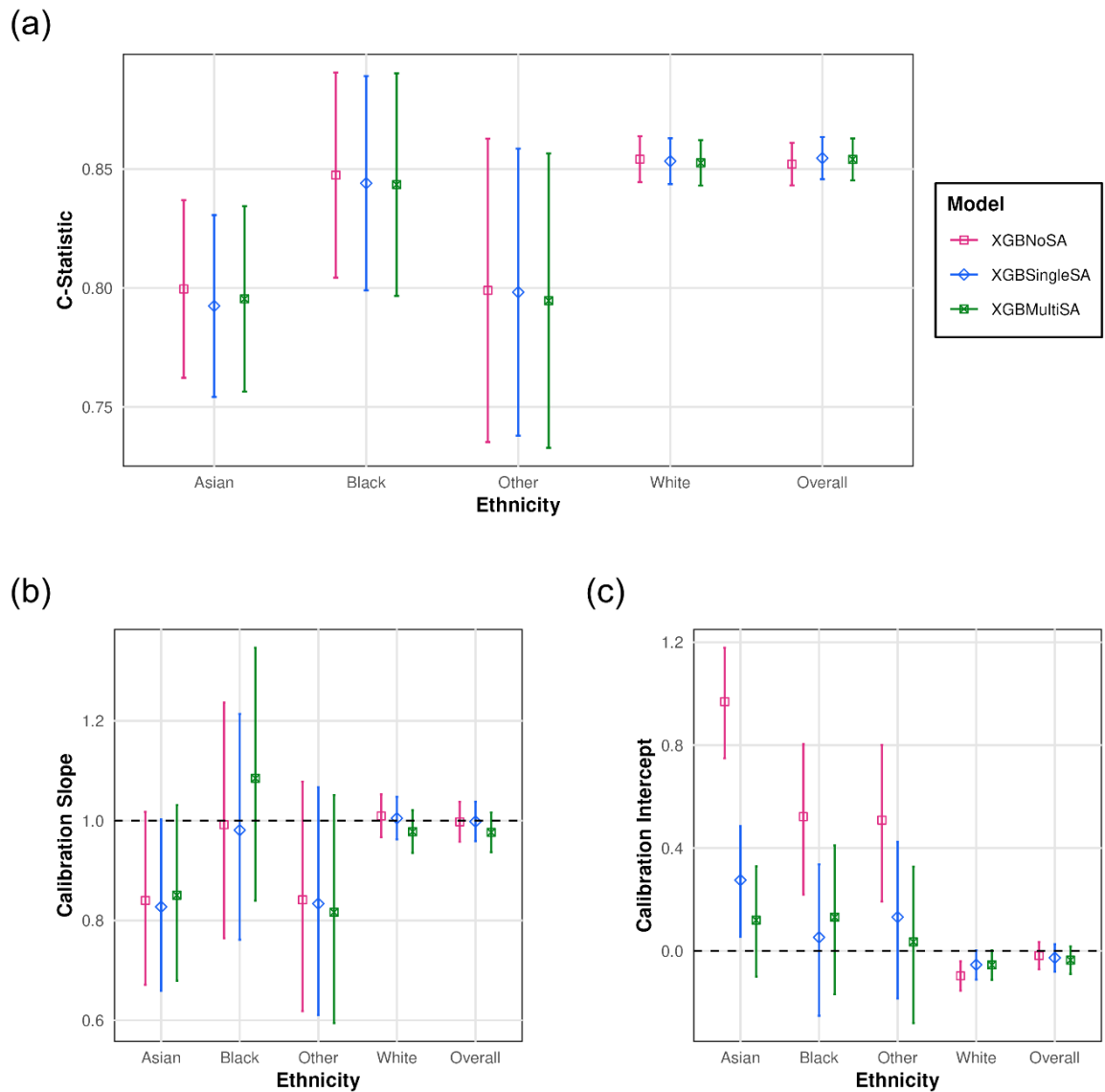

**Supplementary Figure 3:** Subgroup net benefit, in true negatives per 10,000 patients, in 5-year T2 diabetes risk prediction in different ethnicities and overall, for three models: a XGBoost model which does not use ethnicity as a predictor (XGBNoSA), a XGBoost model which does include it (XGBSingleSA) and an ethnicity-personalised ensemble of XGBoost models (XGBMultiSA). A policy of screening no patients (Treat.No.One) is also plotted.

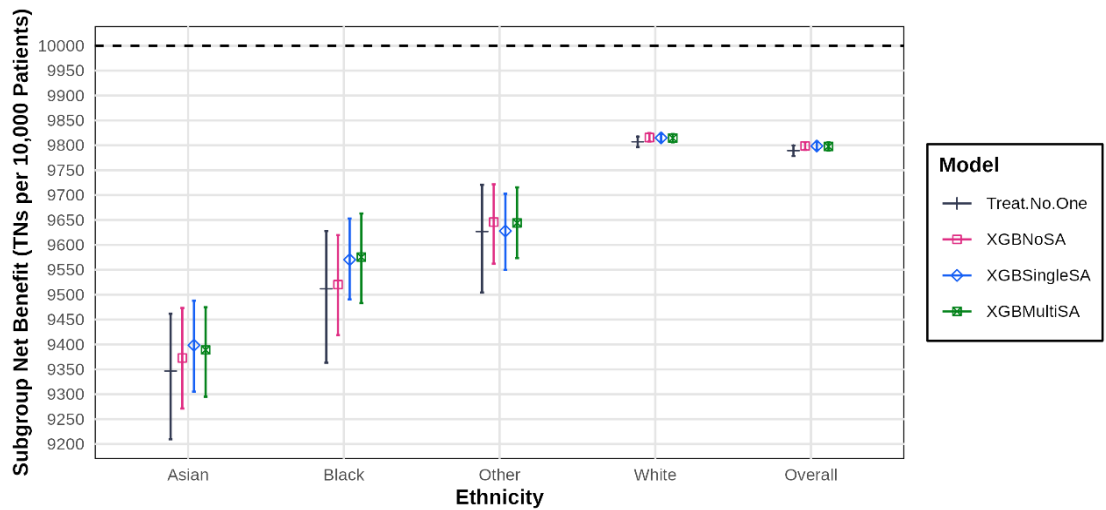

**Supplementary Figure 4:** Net benefit (in true positives per 10,000 patients) across thresholds for the overall (a), Asian (b), Black (c), other ethnicity (d), and white (e) groups for three models: a XGBoost model which does not use ethnicity as a predictor (XGBNoSA), a XGBoost model which does include it (XGBSingleSA) and an ethnicity-personalised ensemble of XGBoost models (XGBMultiSA). The net benefit of treating no patients (Treat.No.One) and all patients (Treat.All) are also plotted. The chosen optimal threshold to calculate the subgroup net benefit (15%) is shown through a vertical orange line.

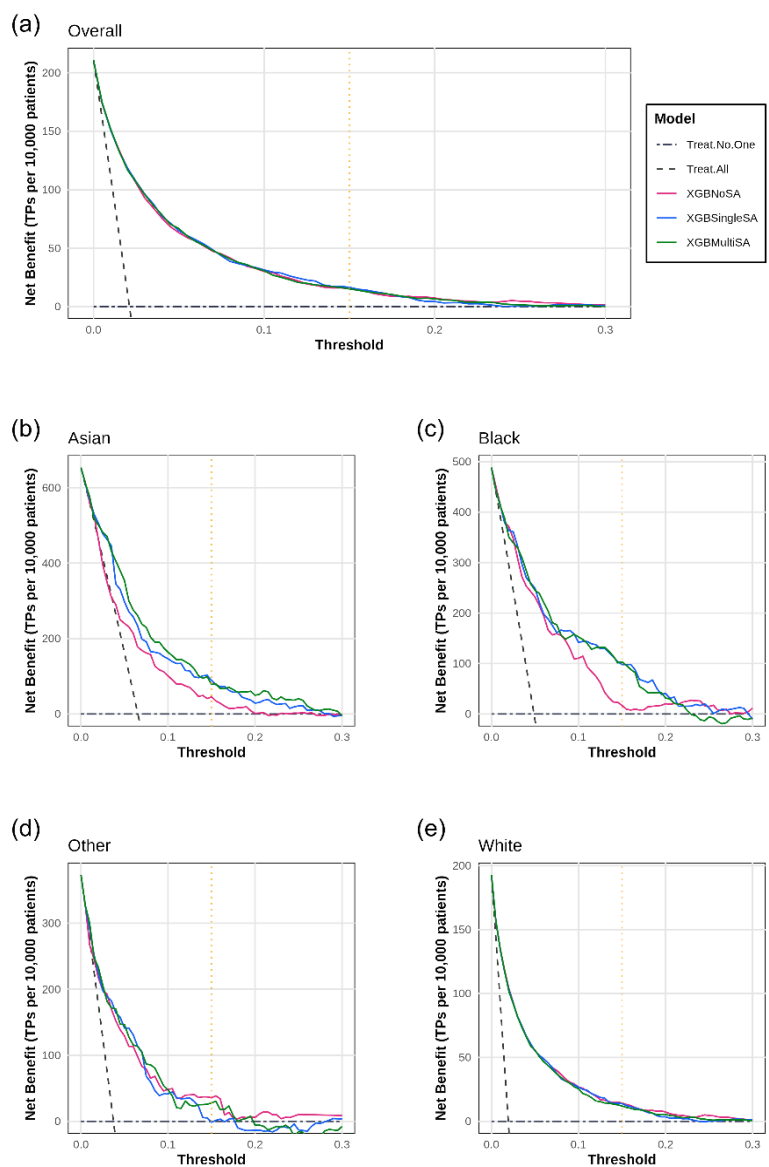

**Supplementary Table 3:** Demographic characteristics of the cohort identified for the lung cancer prediction problem, stratified by Townsend quintile. Townsend was missing for 599 individuals, not appearing on this table. For continuous values, the median and interquartile range are reported, while for categorical values, the count and percentage are reported.

| Characteristic | 1st (Lowest Deprivation), N = 95,394 (20%) | 2nd, N = 96,366 (20%) | 3rd, N = 95,720 (20%) | 4th, N = 96,411 (20%) | 5th (Highest Deprivation), N = 95,998 (20%) |
|---|---|---|---|---|---|
| **Age** | 58 (51, 63) | 59 (51, 63) | 58 (50, 63) | 57 (49, 63) | 56 (48, 62) |
| **Female** | 51,846 (54%) | 52,718 (55%) | 53,057 (55%) | 53,410 (55%) | 51,377 (54%) |
| **Townsend Score** | -4.58 (-5.05, -4.24) | -3.35 (-3.65, -3.06) | -2.15 (-2.48, -1.77) | -0.17 (-0.80, 0.53) | 3.39 (2.26, 4.92) |
| **Ethnicity** | | | | | |
| **Asian** | 1,111 (1.2%) | 1,170 (1.2%) | 1,637 (1.7%) | 3,173 (3.3%) | 4,029 (4.2%) |
| **Black** | 213 (0.2%) | 352 (0.4%) | 635 (0.7%) | 1,600 (1.7%) | 4,967 (5.2%) |
| **Other** | 840 (0.9%) | 944 (1.0%) | 1,149 (1.2%) | 1,860 (1.9%) | 3,796 (4.0%) |
| **White** | 93,127 (98%) | 93,777 (97%) | 92,143 (96%) | 89,527 (93%) | 82,579 (87%) |
| **Unknown** | 103 | 123 | 156 | 251 | 627 |
| **Education** | | | | | |
| **None Reported** | 10,760 (13%) | 13,178 (16%) | 15,296 (18%) | 17,186 (20%) | 24,432 (30%) |
| **GCSE-Equivalent** | 25,737 (31%) | 27,209 (32%) | 26,836 (32%) | 25,187 (30%) | 22,650 (27%) |
| **A Level-Equivalent** | 12,373 (15%) | 11,695 (14%) | 10,891 (13%) | 10,152 (12%) | 8,334 (10%) |
| **University-Equivalent** | 34,612 (41%) | 31,747 (38%) | 30,373 (36%) | 31,936 (38%) | 27,357 (33%) |
| **Unknown** | 11,912 | 12,537 | 12,324 | 11,950 | 13,225 |
| **Non-Lung Cancer in Family** | 87,840 (92%) | 88,177 (92%) | 87,410 (91%) | 87,727 (91%) | 85,227 (89%) |
| **Lung Cancer in Family** | 11,133 (12%) | 11,914 (12%) | 12,827 (13%) | 14,537 (15%) | 19,419 (20%) |
| **Chronic Bronchitis in Family** | 10,836 (11%) | 11,730 (12%) | 12,408 (13%) | 13,555 (14%) | 17,493 (18%) |
| **Smoking Status** | | | | | |
| **Current Smoker** | 5,894 (6.2%) | 6,805 (7.1%) | 8,224 (8.6%) | 11,120 (12%) | 18,097 (19%) |
| **Never Smoked** | 57,763 (61%) | 56,174 (59%) | 53,855 (57%) | 50,946 (53%) | 44,324 (47%) |
| **Previous Smoker** | 31,391 (33%) | 33,008 (34%) | 33,167 (35%) | 33,812 (35%) | 32,549 (34%) |
| **Unknown** | 346 | 379 | 474 | 533 | 1,028 |
| **Taking Statins** | 14,515 (15%) | 15,592 (16%) | 15,846 (17%) | 16,328 (17%) | 19,082 (20%) |
| **Taking Blood Pressure Medication** | 17,827 (19%) | 18,979 (20%) | 19,213 (20%) | 19,566 (20%) | 21,711 (23%) |
| **Emphysema** | 1,056 (1.1%) | 1,186 (1.2%) | 1,485 (1.6%) | 1,838 (1.9%) | 3,122 (3.3%) |
| **Previous Cancer** | 4,361 (4.6%) | 4,338 (4.5%) | 4,081 (4.3%) | 3,674 (3.8%) | 3,259 (3.4%) |
| **6-Year Lung Cancer Incidence** | 236 (0.2%) | 289 (0.3%) | 299 (0.3%) | 393 (0.4%) | 644 (0.7%) |

**Supplementary Figure 5:** Net benefit (in true positives per 10,000 patients) across thresholds for the groups as defined by the quintile of their Townsend score, for three models: a logistic regression model which does not use ethnicity as a predictor (LogNoSA), a logistic regression model which does include it (LogSingleSA) and an ethnicity-personalised ensemble of logistic regression models (LogMultiSA). The net benefit of treating no patients (Treat.No.One) and all patients (Treat.All) are also plotted. The chosen optimal threshold to calculate the subgroup net benefit (1.5%) is shown through a vertical orange line.

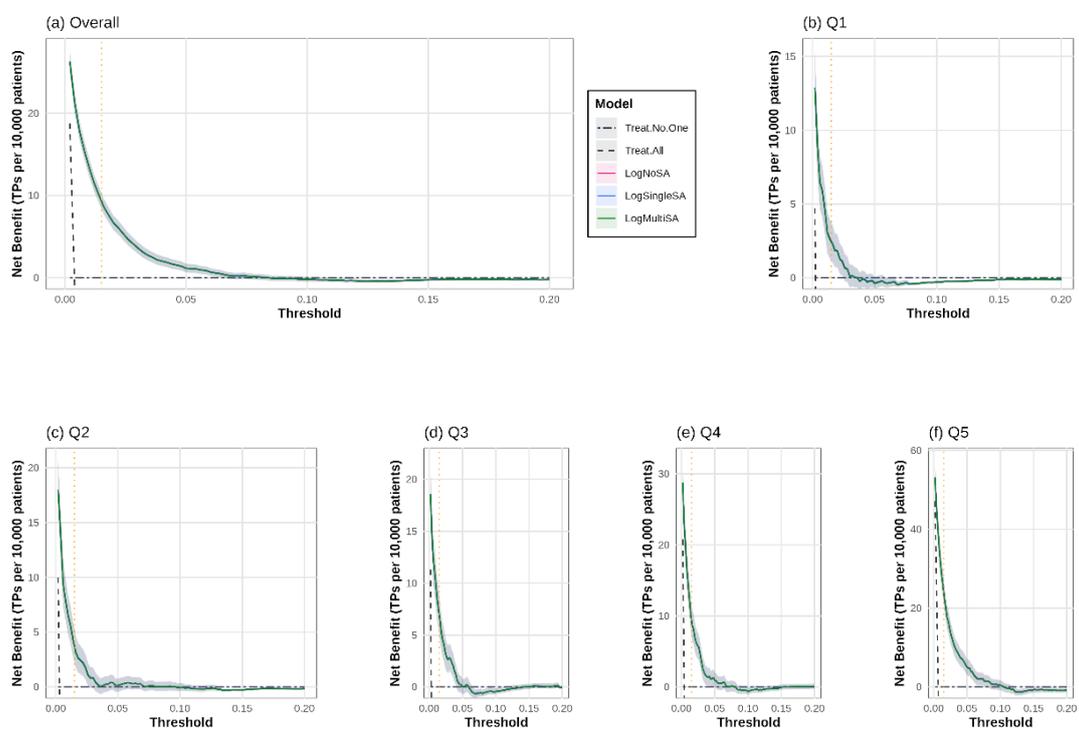

**Supplementary Figure 6:** Performance in 6-year lung cancer screening for different quintiles of Townsend score and overall, for three models: a XGBoost model which does not use the Townsend score as a predictor (XGBNoSA), a XGBoost model which does include it (XGBSingleSA) and a quintile-personalised ensemble of XGBoost models (XGBMultiSA). The C-statistic (a), calibration slope (b) and calibration intercept (c) are plotted with 95% confidence intervals. "Q5" corresponds the most deprived quintile of the population.

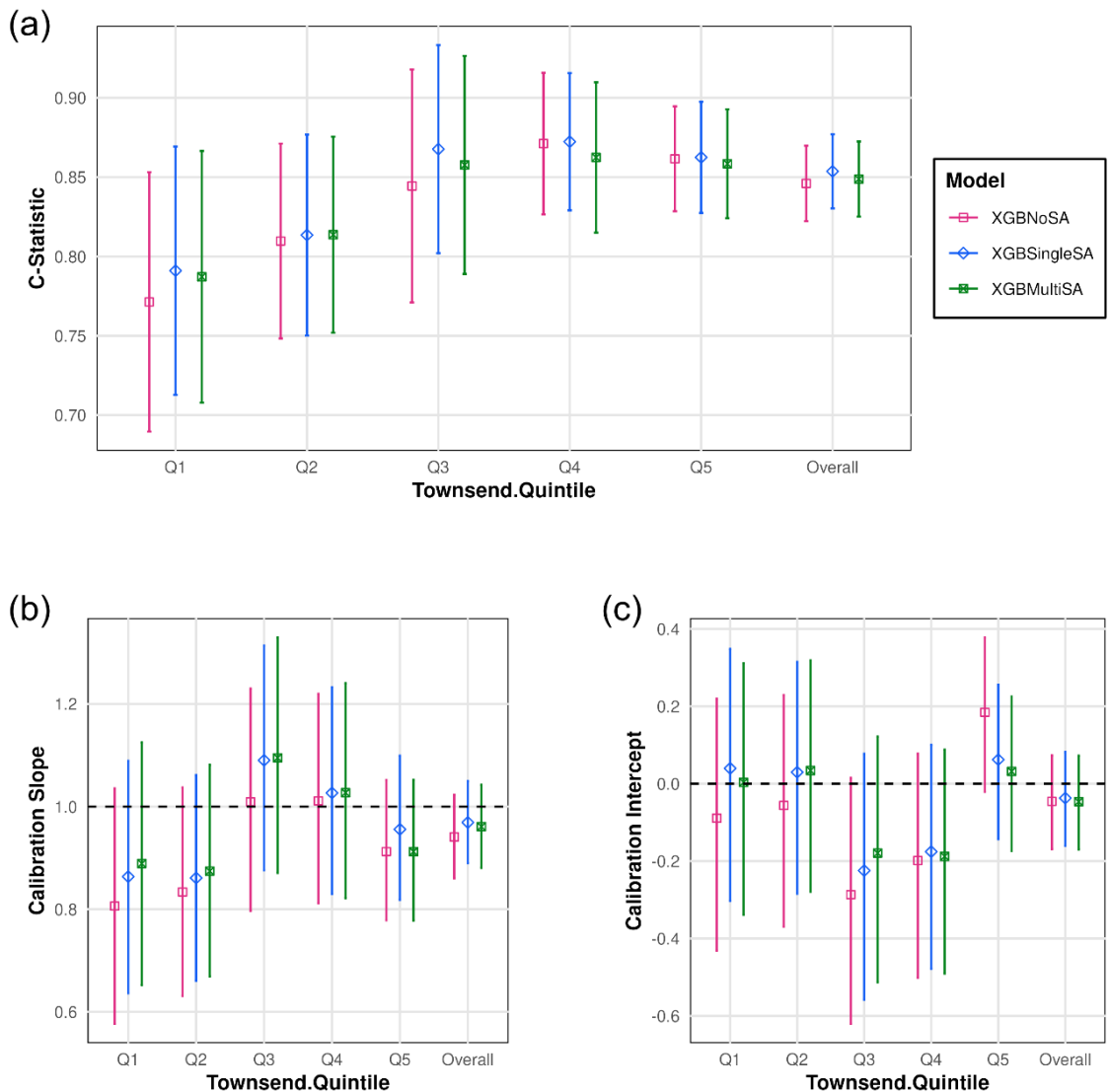

**Supplementary Figure 7:** Subgroup net benefit, in true negatives per 10,000 patients, in 6-year lung cancer screening for different quintiles of Townsend score and overall, for three models: a XGBoost model which does not use Townsend score as a predictor (XGBNoSA), a XGBoost model which does include it (XGBSingleSA) and a quintile-personalised ensemble of XGBoost models (XGBMultiSA). A policy of screening no patients (Treat.No.One) is also plotted.

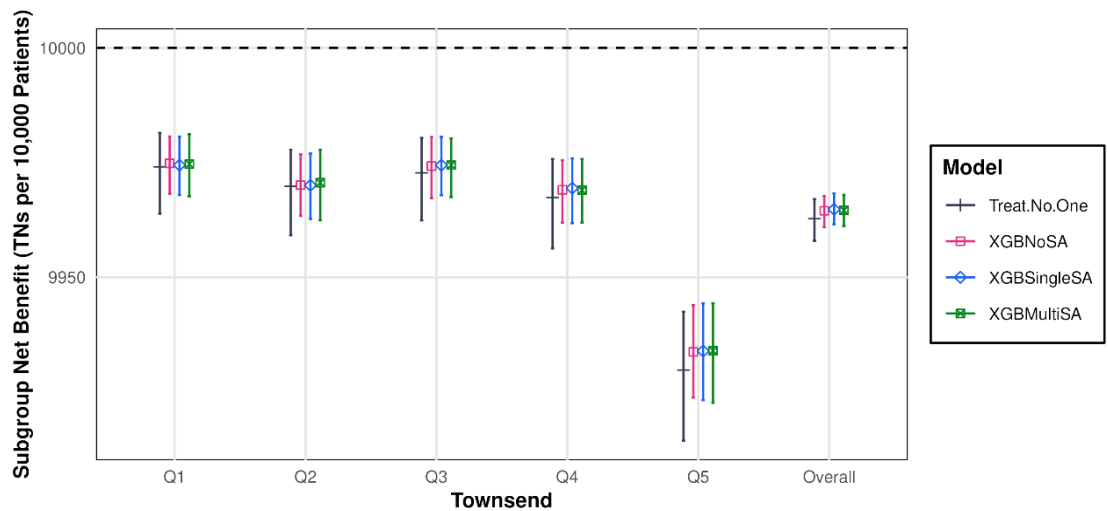

**Supplementary Figure 8:** Net benefit (in true positives per 10,000 patients) across thresholds for the groups as defined by the quintile of their Townsend score, for three models: a XGBoost model which does not use ethnicity as a predictor (XGBNoSA), a XGBoost model which does include it (XGBSingleSA) and an quantile-personalised ensemble of XGBoost models (XGBMultiSA). The net benefit of treating no patients (Treat.No.One) and all patients (Treat.All) are also plotted. The chosen optimal threshold to calculate the subgroup net benefit (1.5%) is shown through a vertical orange line.

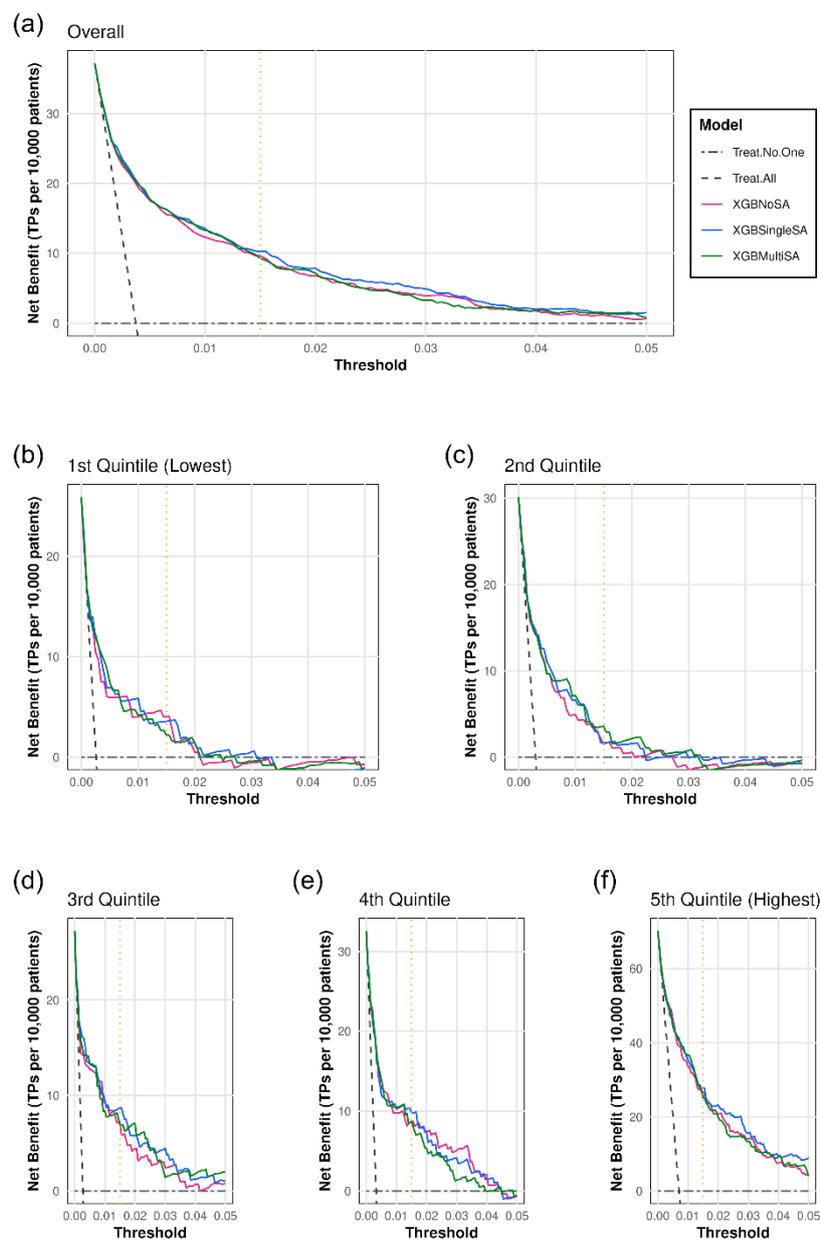

**Supplementary Figure 9:** Trade-off between the subgroup net benefit in the most deprived quintile and overall, shown through the identified set of thresholds that are Pareto optimal, for three models: a logistic regression model which does not use Townsend score as a predictor (XGBNoSA), a logistic regression model which does include it (XGBSingleSA) and a quintile-personalised ensemble of logistic regression models (XGBMultiSA). The Pareto front was found in the training dataset, without cross-validation, (a), and the identified choice of thresholds was plotted in the validation dataset (b). The policy of screening no patients (Treat No One) is also shown.

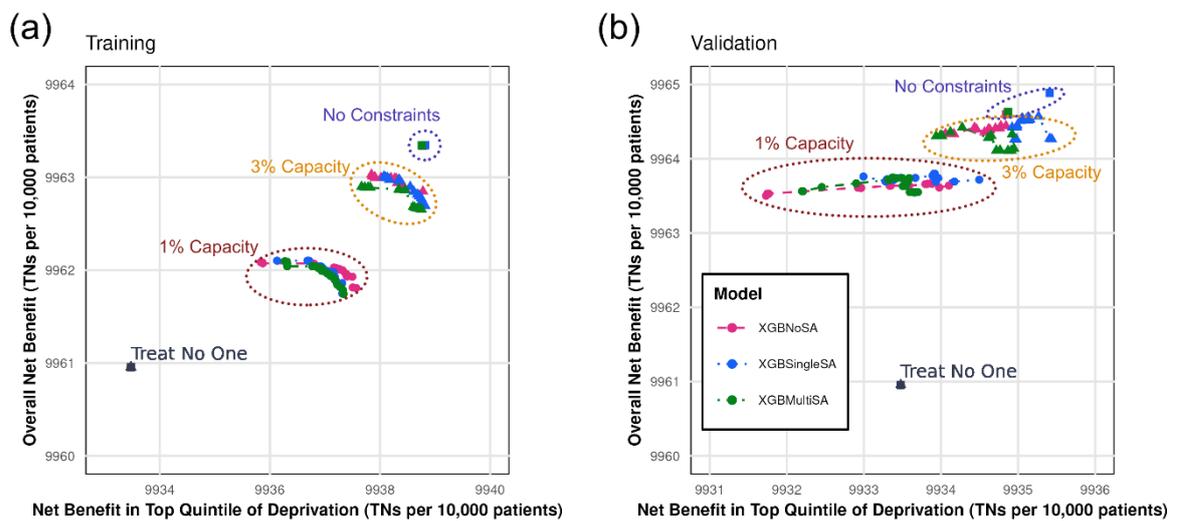